\documentclass[conference]{IEEEtran}
\IEEEoverridecommandlockouts
\usepackage{cite}
\usepackage{amsmath,amssymb,amsfonts}
\usepackage{algorithmic}
\usepackage{graphicx}
\usepackage{textcomp}
\def\BibTeX{{\rm B\kern-.05em{\sc i\kern-.025em b}\kern-.08em
    T\kern-.1667em\lower.7ex\hbox{E}\kern-.125emX}}
\begin{document}

%\title{SmartCapAC: Capability-based Access Control For the Internet of Things (IoTs) Using Blockchain}

\title{BlendCAC: A BLockchain-ENabled Decentralized Capability-based Access Control for IoTs}

\author{
\IEEEauthorblockN{Ronghua Xu, Yu Chen}
\IEEEauthorblockA{%\textit{Dept. of Electrical \& Computer Engineering} \\
\textit{Binghamton University, SUNY} \\ Binghamton, NY 13902, USA \\
\{rxu22, ychen\}@binghamton.edu}
\and
\IEEEauthorblockN{Erik Blasch}
\IEEEauthorblockA{\textit{US Air Force Research Laboratory} \\
Rome, NY 13441, USA \\
erik.blasch@gmail.com}
\and
\IEEEauthorblockN{Genshe Chen}
\IEEEauthorblockA{\textit{Intelligent Fusion Technology, Inc.} \\
Germantown, MD 20876, USA \\
gchen@intfusiontech.com}
}

\maketitle

\begin{abstract}
The prevalence of Internet of Things (IoTs) allows heterogeneous embedded smart devices to collaboratively provide smart services with or without human intervention. While leveraging the large-scale IoT-based applications like Smart Gird or Smart Cities, IoTs also incur more concerns on privacy and security. Among the top security challenges that IoTs face, access authorization is critical in resource sharing and information protection. One of the weaknesses in today's access control (AC) is the centralized authorization server, which can be the performance bottleneck or the single point of failure. In this paper, BlendCAC, a blockchain-enabled decentralized capability-based AC is proposed for the security of IoTs. The BlendCAC aims at an effective access control processes to devices, services and information in large scale IoT systems. Based on the blockchain network, a capability delegation mechanism is suggested  for access permission propagation. A robust identity-based capability token management strategy is proposed, which takes advantage of smart contract for registering, propagation and revocation of the access authorization. In the proposed BlendCAC scheme, IoT devices are their own master to control their resources instead of being supervised by a centralized authority. Implemented and tested on a Raspberry Pi device and on a local private blockchain network, our experimental results demonstrate the feasibility of the proposed BlendCAC approach to offer a decentralized, scalable, lightweight and fine-grained AC solution to IoT systems.
\end{abstract}


%===================================== 1. Introduction ============================================

\section{Introduction}
\label{sec:intro}  % \label{} allows reference to this section
Thanks to the proliferation of the Internet of Things (IoTs), more and more smart devices are connected to the Internet at an unprecedented scale. The prevalence of the IoTs changes human activities by pervasively providing applications and services revolutionizing transportation, healthcare, industrial automation, and emergency response \cite{al2015internet}. These capabilities enable both situational awareness (SAW) and information fusion based on uninterrupted, ubiquitous, real-time data flows, which provide essential contextual information for delay-sensitive, mission-critical applications \cite{snidaro2016context, blasch2017panel,chen2016dynamic,chen2016smart}.

While benefiting from the large-scale applications like Smart Gird and Smart Cities, highly connective smart IoT devices with insufficient security enforcement incur more concerns on security and privacy. Security issues, such as privacy, authentication, access control, system configuration, information storage and management, are among the top challenges in today's IoT environment \cite{alaba2017internet}.

One of the top security concerns that IoTs face is access control (AC). Access authorization is particularly critical in resource and information protection. Conventional AC approaches, such as the Access Control List (ACL), the Role-based Access Control (RBAC) and the Attribute-based Access Control (ABAC), have been widely used in IT systems. However, they are not able to provide a scalable, manageable and efficient mechanism to meet new challenges raised by IoT networks:

% \begin{itemize} define number item list
\begin{enumerate}
\item \textit{Scalability}: The fast growing number of devices and services also pose increasing management workload to AC systems. AC strategies are expected to be able to handle the scalability problem resulting from the large scale IoT networks.

\item \textit{Heterogeneity}: IoT systems normally integrate heterogeneous computing nodes and other physical devices with variant underlying technologies or in different application domains, and each domain or platform has its own specific requirements for identity authentication and authorization policy enforcement.

\item \textit{Causality}: The traditional RBAC and ABAC systems envisage planned and long-lived patterns, while the IoT world is mainly characterized by short-lived, with causal and/or spontaneous interactions \cite{gusmeroli2013capability}, in which the AC scheme is required to dynamically respond.

\item \textit{Lightweight}: IoT devices are usually resource-constrained, which implies the lack of capability to support computing intensive tasks or large storage required applications. meanwhile, the smart devices connect to each other by low power and lossy networks. Consequently the AC protocols should be lightweight and not impose significant overhead on devices and communication networks.
\end{enumerate}

The extraordinary large number of devices with heterogeneity and dynamics necessitate a more scalable, flexible and lightweight AC mechanism for IoT networks. In addition, a majority of the AC solutions rely on centralized authorities. Although the delegation mechanism helps migrate certain intelligence from the centralized cloud server to a near-site fog or edge network, the power of policy decision making and identity management is exclusively located in the cloud center. IoT networks need a new AC framework that provides decentralized authentication and authorization schemes in trustless application network environments, such that intelligence could be diffused among large number of distributed edge devices. 

While being well-known as the fundamental protocol of Bitcoin \cite{nakamoto2008bitcoin}, the first digital currency, the blockchain protocol has been recognized as the potential to revolutionize the fundamentals of IT technology because of its many attractive features and characteristics such as supporting decentralization and anonymity maintenance \cite{crosby2016blockchain}. In this paper, a \textit{BLockchain-ENabled, Decentralized, Capability-based Access Control} (BlendCAC) scheme is proposed to enhance the security of IoTs. It provides a scalable, fine-grained, and lightweight AC solution to protect smart devices, services and information in IoT networks. An identity-based capability token management strategy is presented and the authorization delegation mechanism is illustrated. A capability-based access validation process is implemented on service providers that integrate situational awareness (SAW) and customized contextualized conditions. Experimental results demonstrate the feasibility and effectiveness of the proposed AC scheme. 

The major contributions of this work are:

\begin{itemize}
\item[1)] A complete architecture of capability-based authorization is proposed, which includes capability management, delegation propagation, and access right validation.

\item[2)] A concept-proof prototype based on smart contracts is implemented and deployed on a local private blockchain network.

\item[3)] A comprehensive experimental study has been conducted that compares the proposed BlendCAC scheme with the well-known RBAC and ABAC models. The experimental results validate the feasibility of the BlendCAC scheme in IoT environments without introducing significant overhead.
\end{itemize}

The remainder of this paper is organized as follows: Section \ref{sec:related} analyzes and reviews the state of the art research in access control for IoT systems. Section \ref{sec:BlendCAC} illustrates the details of the proposed BlendCAC system. Beside the implementation of the proof-of-concept prototype, Section \ref{sec:experiment} reports an extensive experimental study using test scenarios that are run on both resource-constrained and non-resource-constrained devices. Finally, a summary wraps up this paper in Section \ref{sec:conclusion}.

% =========================================== 2.related work =============================================
\section{Background Knowledge and Related Work}
\label{sec:related}  % \label{} allows reference to this section

\subsection{Access Control in IoTs}

Technologies for authentication and authorization of access to certain resources or services are among the main elements to protect the security and privacy for IoT devices \cite{ouaddah2017access}. As a fundamental mechanism to enable security in computer systems, AC is the process that decides who is authorized to have what communication rights on which objects with respect to some security models and policies \cite{gong1989secure}. An effective AC system should satisfy the main security requirements, such as confidentiality, integrity, and availability. However, recently raised security and privacy issues in the era of IoTs require that AC systems should be built on principals of high scalability, flexibility, lightweight and causality.

There are various AC methods and solutions with different objectives proposed to address security challenges in IoTs. The Role-Based Access Control (RBAC) model \cite{sandhu1996role} provides a framework that species user access authorization to resources based on roles, and supports principals such as least privilege, partition of administrative functions and separation of duties \cite{samarati2000access}. However, a pure RBAC model presents a role explosion problem,  which is inappropriate to implement security policies that require interpreting complex and ambiguous IoT scenarios. The RBAC model implemented on IoTs adopts a Web of Things (WoTs) approach to implement AC policies on the smart objects via the web service \cite{de2008socrades,spiess2009soa}, and the RBAC model was extended by introducing context constraints to consider contextual awareness in AC decisions \cite{zhang2010extended}. However, those proposals are not able to clearly specify the fine-grained AC on variant resources or services, like the mapping of the role notion and device-to-device communication.

To address the weaknesses of RBAC model in a highly distributed network environment, an Attribute-based Access Control (ABAC) \cite{yuan2005attributed, smari2014extended} is introduced in IoT networks to reduce the number of rules resulting from role explosion. In ABAC the AC policies are defined through directly associating attributes with subjects. An efficient authentication and ABAC based authorization scheme for the perception layer of IoTs have been proposed \cite{ye2014efficient}. Based on user attribute certificates, an access right is granted by AC authority to ensure fine-grained access control. However, specifying a consistent definition of the attributes within a domain or across different domains could significantly increase effort and complexity on policy management as the number of devices grow, and hence, the attribute-based proposal is not suitable for large scale distributed IoT networks.

Due to drawbacks that exist in traditional access control models such as RBAC and ABAC, the requirements imposed by IoT scenarios cannot be satisfied. Given many great advantages from an IoT perspective, such as scalability, flexibility, distributed, user-driven, IoT systems can support delegation and revocation\cite{ouaddah2017access}. Capability-based access control approaches have been considered a promising solution to IoTs. The Access Control Matrix (ACM) model represents a good conceptualization of authorizations by providing a framework for describing discretionary access control (DAC) \cite{samarati2000access}. As two implementations of ACM, Access Control List (ACL) and Capability are widely used in authorization system. In the ACL model, each object is associated with an access control list that saves the subjects and their access rights for the objects. 

The ACL is a centralized approach to support administrative activities with better traceability by implementing AC strategies on cloud servers \cite{liu2014information}. However, as the number of subjects and resources increases, confused duty problems are identified in ACL and access rules become much more complex to manage. Due to the centralized management property, ACL cannot provide multiple levels of granularity, is not scalable and is vulnerable to single point of failure. Meanwhile, in the capability model, each subject is associated with a capability list that represents its access rights to all concerned objects. The CapAC has been implemented in many large scale IoT-based projects, like IoT@Work \cite{gusmeroli2012iot}.

Although capability-based methods have been used as a feature in many access control solutions for the IoT-based applications, applying the original concept of capability-based access control model in IoT network has raised several issues, like capability propagation and revocation \cite{gong1989secure}. To tackle these challenges, a Secure Identity-Based Capability (SICAP) System is proposed, which enables the monitoring, mediating, and recording of capability propagations to enforce security policies as well as achieving rapid revocation capability by using an exception list \cite{gong1989secure}. However, the centralized access control server (ACS) becomes the performance bottleneck of the system, and the author didn't provide a clear illustration on security policy used in capability generation and propagation, neither was the context information in making authorization decision considered. 

To enable contextual awareness in federated IoTs, an authorization delegation method is proposed based on a Capability-based Context-Aware Access Control (CCAAC) model \cite{anggorojati2012capability}. By introducing a delegation mechanism to capability generation and propagation process, the CCAAC model shows great advantages to address scalability and heterogeneity issues in IoT networks. Given the requirement that a prior knowledge of the trust relationship among domains in federated IoTs must be established, however, the proposed approach is not suitable universally for all IoT application scenarios. Inspired by the SUN DIGITAL ECOSYSTEM ENVIRONMENT project \cite{skinner2009cyber}, a capability-based access control (CapAC) model was proposed that adopted a centralized approach for managing access control policy \cite{gusmeroli2013capability}. However, the proposed CapAC scheme depended on a centralized authority and did not consider the lightweight requirement at the smart device side. To address the limitations in CapAC, a Distributed Capability-based Access Control (DCapAC) model was proposed, which was directly deployed on resource-constrained devices \cite{hernandez2013distributed,hernandez2016dcapbac}. The DCapAC allows smart devices to autonomously make decisions on access rights based on authorization policy, and it shows advantages in scalability and interoperability. However, capability revocation management and delegation were not discussed, neither were the granularity and context-awareness considered.

%--------------------2.2 Blockchain and Smart contract introduction --------------------
\subsection{Blockchain and Smart Contract}

The blockchain is the fundamental framework of Bitcoin \cite{nakamoto2008bitcoin}, which was introduced by Nakamoto in 2008. The blockchain is the public ledger that allows the data be recorded, stored and updated distributively. By its nature, the blockchain is a decentralized architecture such that does not rely on a centralized authority anymore. The transactions are approved and recorded in blocks by miners, and the blocks are appended to the blockchain in a chronological order. Blockchain uses consensus mechanism to maintain the sanctity of the data recorded on the blocks. Thanks to the “trustless” proof mechanism enforced through mining task on miners across network, users can trust the system of the public ledger stored worldwide on many different decentralized nodes maintained by ''miner-accountants,'' as opposed to having to establish and maintain trust with the transaction counter-party or a third-party intermediary \cite{swan2015blockchain}. Blockchain is the ideal architecture to ensure distributed  transactions between all participants in a trustless environment.

Because of many attractive characteristics, blockchain technology has been investigated to offer a decentralized AC scheme in trustless network environments. A blockchain based AC is proposed to publish AC policy and to allow distributed transfer access right among users on bitcoin network \cite{maesa2017blockchain}. The proposal allows distributed auditability, preventing a third party from fraudulently denying the rights granted by an enforceable policy. However, the solution still rely on an external centralized policy database to fetch access right given the links stored in the blockchain, and the experimental results are not provided. Based on blockchain technology, FairAccess is proposed to offer a fully decentralized pseudonymous and privacy preserving authorization management framework for IoTs \cite{ouaddah2016fairaccess}. In FairAccess, AC policies are enclosed in new types of transactions that are used to grant, get, delegate, and revoke access. However, the scripting language used in Bitcoin allows to transcode two types of AC policies, so that the proposed framework cannot support more complex and granular access control model.

Blockchain has shown its success in decentralization of currency and payments, like Bitcoin. Currently designing a programmable money and contracts, which support variety of customized transaction types become a trend to extend blockchain application beyond cryptocurrency. Smart contract, which emerges from the smart property, is a method of using blockchain to achieve agreement among parties, as opposed to rely on third parties for maintaining a trust relationship. By using cryptographic and other security mechanisms, smart contract combines protocols with user interfaces to formalize and secure relationships over computer networks\cite{szabo1997formalizing}. Smart contract is essential a collection of defined instructions and data that have been recorded at a specific address of blockchain. Through encapsulating operational logic as a bytecode and performing Turing complete computation on distributed miners, a smart contract allows user to transcode more complex business models as new types of transactions on a blockchain network. Smart contract provides a promising solution to implement more flexible and fine-grained AC models on blockchain networks. 

\section{BlendCAC: a BLockchain-ENabled Decentralized CapAC}
\label{sec:BlendCAC}  % \label{} allows reference to this section

In order to address the issues discussed in Section \ref{sec:related}, the authors recently proposed a Federated Capability-based Access Control model (FedCAC) \cite{xu2017federate}. FedCAC addresses scalability, granularity, and dynamicity challenges in access control strategy for IoTs. Through delegating part of the identify authentication and authorization task to domain delegator, workload of the centralized policy decision making center (PDC) is reduced. Migrating some processing validation tasks to local devices helps the FedCAC to be lighter and context-awareness enabled. Involving smart objects in access right authorization process allows device-to-device communication, which implies better scalability and interoperability in IoT network environment.
However, it is essentially still a centralized AC scheme, such that weaknesses include being the single-point of failure and performance bottleneck, are still not solved.  

Inspired by the smart contract and blockchain technology, a decentralized capability-based access control framework for IoTs, called BlendCAC, is proposed in this paper, and a prototype of proposal has been implemented in a physical IoT network environment to verify the efficiency and effectiveness. The next subsection provides a comprehensive system design of BlendCAC framework. Unlike the approaches discussed above, BlendCAC, effectively provides scalability, granularity, and dynamicity of AC strategies for IoTs. Through encapsulating AC policies into a Smart Contract, which is deployed across the blockchain network, users are the  master of their own data or devices instead of being supervised or controlled by a third party authority. Enforcing authorization and access right verification among large number of distributed edge devices allows more coordination on edge networks.

%In most IoT based systems, data processing and security enforcement are deployed on centralized cloud centers where abundant computing and storage resources are available. Consequently, all access right requests from devices are transmitted to remote servers for authentication and authorization. Such a centralized network architecture is not scalable for large scale of IoT networks, and latency are too large to many mission-critical applications. To meet the requirements of real-time processing and instant decision making in IoT-based applications, researchers has proposed an online, uninterrupted smart surveillance system by leveraging the fog computing paradigm \cite{chen2016dynamic, chen2017enabling, chen2016smart}. Given a network with established trust relationship, the proposed hierarchical architecture allows to delegate part of the AC policies from the server to some near-site fog computing nodes. However, users and devices still rely on centralized authority, either on cloud or fog nodes, to manage their data and resources. The blockchain, which allows the disintermediation and decentralization of all transactions in a trustless network \cite{swan2015blockchain}, becomes a promising solution to address AC issues in IoT systems.

 %------------- 3.1-System design of BlendCAC ---------------------
\subsection{System Architecture of BlendCAC}

Figure \ref{fig:1-BlendCAC} illustrates the proposed BlendCAC system architecture, which intends to function in a scenario including two isolated IoT-based service domains without pre-establishing a trust relationship. In each domain, the domain owner, who has the ownership of devices, enforces predefined security policies to manage domain related devices and services. Operation and communication mode is listed as follow:

%Fig1_System_architecture_BlendCAC
\begin{figure} [t]
\begin{center}
\begin{tabular}{c}
\includegraphics[height=4cm]{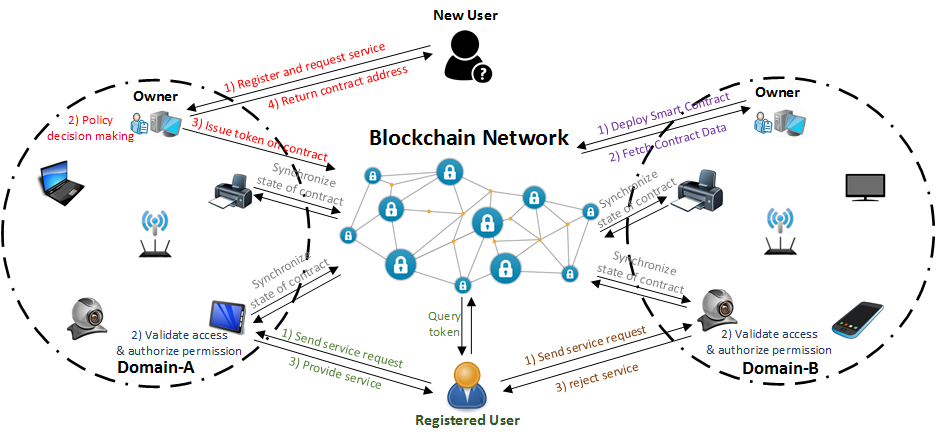}
\end{tabular}
\end{center}
\caption[example] {\label{fig:1-BlendCAC} Illustration of the BlendCAC System Architecture.}
\vspace{-10pt}
\end{figure}

\begin{enumerate}
\item \emph{Registration}: All entities must create at least one main account defined by a pair of keys to join the blockchain network. Each account is indexed by its address that is derived from his/her own public key. In our scenario, the domain owner maintains a local profile database that provides identity authentication and management. Once the identity information related to users or IoT devices is verified, the profile of each registered entity is created by using his/her address for authentication process when an access right request happens. As a result, the domain owners are able to define authorization policies and perform decision-making to directly control their own devices or resources instead of depending on third parties.  

\item \emph{Smart Contract Deployment}: A smart contract, which manages capability tokens, must be developed and deployed on the blockchain network by the domain owner. Thanks to cryptographic and security mechanisms provided by blockchain network, smart contracts can secure any algorithmically specifiable protocols and relationships from malicious interference by third parties under trustless network environment. After synchronizing the blochchain data, all nodes could access all transactions and recent state of each smart contract by referring local chain data. Each node interacts with the smart contract through the provided contract address and the RPC interface.

\item \emph{Capability Propagation}: To successfully access services or resources at service providers, an entity initially sends an access right request to the domain owner to get a capability token. Given the registered entity information established in the profile database, a policy decision making module evaluates the access request by enforcing the predefined authorization policies. If the access request is granted, the domain owner issues the capability token encoded the access right, then launches a transaction to update the token data in the smart contract. After the transaction has been approved and recorded in a new block, the domain owner notifies the entity with a smart contract address for the querying token data. Otherwise, the access right request is rejected. 

\item \emph{Authorization Validation}: The authorization validation process is performed at the local service providers on receiving a service request from the entity. Through regularly synchronizing the local chain data with the blockchain network, a domain owner just simply checks the current state of the contract in the local chain to get a capability token associated with the entity's address. Considering the capability token validation and access authorization process result, if the access right policies and conditional constraints are satisfied, the service provider grants the access request and offers services to the requester. Otherwise, the service request is denied.
\end{enumerate}
%\vspace{1 pt}

To enable a scalable, distributed and fine-grained AC scheme for IoT networks, the proposed BlendCAC is focused on three issues: the identity-based capability management, the access right authorization and the privilege delegation.

%------------------------ 3.2-Identity based Capability Management ------------------------------------
\subsection{Capability Token Structure}
In the BlendCAC system, the entities are categorized as subjects and objects. \textit{Subjects} are defined as entities who request a service from the service providers, while \textit{objects} are referred to entities who offer the resources or services. Entities could be either human beings or smart devices. In the profile database, all registered entities are associated with a globally unique Virtual Identity (VID), which is used as the prime key for identifying entities' profile information. As each entity has at least one main account indexed by its address in the blockchain network, the blockchain is used to represent the VID for profiling register entities.

In general, the capability specifies which subject can access resources of a target object by associating subject, object, actions and condition constraints. The identity-based capability structure is defined as follows:
\begin{equation}
\label{eq:ICap}
ICap = f(VID_{S}, VID_{O}, D, AR, C)
\end{equation}

\noindent{where the parameters are:}
\begin{itemize}
\item $f$: a one-way hash mapping function; 
\item $VID_{S}$: the virtual ID of a subject that requests an access to a service or resource;
\item $VID_{O}$: the virtual ID of an object that provides a service or resource;
\item $D$: a set of delegation right, e.g. delegatee and delegation depth;
\item $AR$: a set of access right for actions, e.g. read, write, execute; and
\item $C$: a set of context awareness information, such as time, location.
\end{itemize}

In the BlendCAC system, an $AR$ is defined as the access right set. For example, the $AR$ can be $\{Read\}$, $\{Write\}$, $\{Read; Write\}$, or $\{NULL\}$. If $AR=\{NULL\}$, the operation conducted on the resource is not allowed. $C$ is defined as a context constraints set, like $C=\{C1, C2\}$ or $C=\{NULL\}$. If $C=\{NULL\}$, no context constraint is considered in the access right validation process. The delegation privilege is defined in set $D$ including \emph{delegatee} and \emph{delegation depth}. The \emph{delegatee} is defined as an array to queue the delegated entity's address, and the \emph{delegation depth} specifies the maximum delegation operations.

%------------------------ 3.3-Authorization and Delegation ------------------------------------
\subsection{Capability-based Access Right Authorization}

\begin{figure} [t]
\begin{center}
\begin{tabular}{c}
\includegraphics[height=5cm]{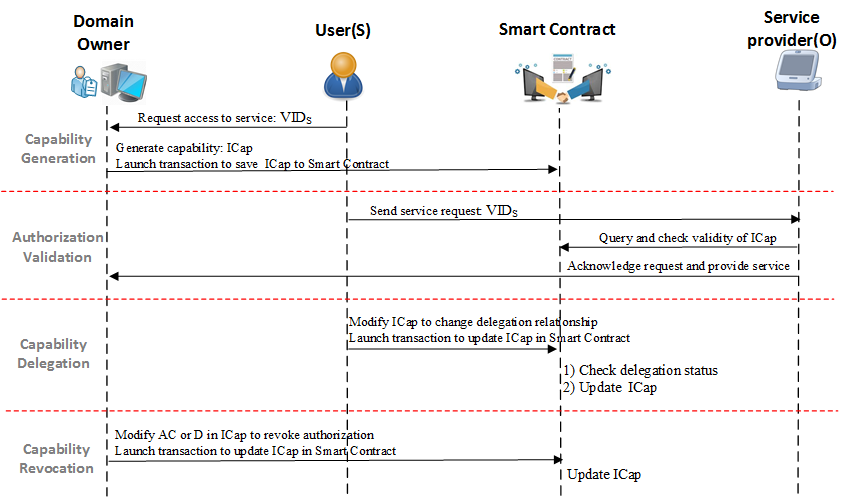}
\end{tabular}
\end{center}
\caption[example] { \label{fig:2-CapAuthorization} Flowchart of the Capability-based Access Right Authorization.}
\vspace{-10pt}
\end{figure}

Given the defined BlendCAC structure, the capability token structure and the related operations are transcoded to a smart contract and deployed on the blockchain network, while the access right authorization is implemented as a policy-based decision making service running by the domain owner. As shown by Fig. \ref{fig:2-CapAuthorization}, a comprehensive capability-based access right authorization procedure consists of four steps: capability generation, access right validation, capability delegation and revocation.

%Fig2_Capability_authorization_process

\begin{enumerate}
\item \emph{Capability Generation}: As one type of meta data to represent the access right, the capability $ICap$ could be generated by associating a VID with an AR, thus the $ICap$ has the identified property to prevent forgery. After receiving access request from user, the domain owner generates capability token based on predefined access control policy, and launches transactions to save a new token data to a smart contract. A large number of $ICap$ is grouped into the capability pools on smart contract, which could be proofed and synchronized among the nodes across the blockchain network.

\item \emph{Access Right Validation}: After receiving the service request from a subject, the service provider first fetches the capability token from the smart contract by using the subject's address, then makes decision whether or not to grant an access to the service according to the local access control policy. Implementing access right validation at the local service provider allows smart objects to be involved in the AC decision making task, which is suitable to offer a flexible and fine-grained AC service in IoT networks.

\item \emph{Capability Delegation}: In our proposal, the capability delegation mechanism is implemented by configuring delegation set $D$ in the $ICap$. As the smart contract has received an update token transaction from the user, it checks the set $D$ to validate the delegation right. If the value of \emph{delegation depth} is more than the count of elements in \emph{delegatee}, the user could delegate his/her capability token to other entity by appending the target address to \emph{delegatee}. Otherwise, the capability delegation request is rejected. For each successful delegation transaction, the length of queue \emph{delegatee} increases by $1$ until reaches the limitation defined by the delegation depth.   

\item \emph{Capability Revocation}: The capability revocation considers two scenarios: delegation right $D$ revocation and  $ICap$ revocation. In our proposal, only the domain owner is allowed to perform revocation operation on a smart contract. In the delegation right revocation process, the domain owner could remove the addresses from \emph{delegatee} to revoke the delegated access right for certain specific entities. In case of $ICap$ revocation, through changing \emph{delegation depth} to zero or clearing the $AR$,  the capability token becomes unavailable to all associated entities.
\end{enumerate}

% ============================================== 4.Implementation and Experiment =========================================
\section{Implementation and Experimental Results}
\label{sec:experiment}  % \label{} allows reference to this section

A concept-proof prototype system has been implemented on a real private Ethereum blockchain network environment. Compared with other open blockchain platform, like Bitcoin and Hyperledger, Ethereum has a more matured ecosystem and is designed to be more adaptable and flexible for development of smart contract and business logic \cite{ehtereum}. The proposed BlendCAC model has been transcoded to a smart contract using Solidity \cite{solidity}, which is a contract-oriented, high-level language for implementing smart contracts. The access authorization and validation policy is enforced as a web service application based on the Flask framework \cite{flask} using Python. The Flask is a micro-framework for Python based on Werkzeug, Jinja 2 and good intentions. The lightweight and extensible micro architectures make the Flask a preferable web solution on resource constrained IoT devices. 

In order to evaluate the performance and the overhead of our BlendCAC scheme, two benchmark models, RBAC and ABAC, are also implemented on the experimental web service system. All transcoded access control models have the similar data structure in smart contract except authorization representation. In RBAC based smart contract, authorization is defined as the approach to bridge the relationship between user and permission, while RBAC based smart contract uses user's attributes as representative format for authorization. Both the RBAC and ABAC need a local database, either to maintain the user-role-permission or to manage the attribute-permission policy for authorization validation process, the profiles and policy rules management are developed by using an embedded SQL database engine, called SQLite\cite{sqlite}. The lower memory and computation cost make the SQLite an ideal database solution to resource constrained system, like Raspberry Pi.

%------------------------------ Environmental Setup -------------------------------------
\subsection{Environmental Setup}
The mining task is performed on a system with stronger computing power, like a laptop or a desktop. Two miners are deployed on a laptop, of which the configuration is as follows: the processor is 2.3 GHz Intel Core i7 (8 cores), the RAM memory is 16 GB and the operating system is Ubuntu 16.04. And other four miners are distributed to four desktops which are empowered with the Ubuntu 16.04 OS, 3 GHz Intel Core TM (2 cores) processor and 4 GB memory. Each miner uses two CPU cores for mining. The IoT devices are two Raspberry PI 3 Model B with the configuration as follows: 1.2GHz 64-bit quad-core ARMv8 CPU, the memory is 1GB LPDDR2-900 SDRAM and the operation system is Raspbian based on the Linux kernel. Unfortunately, the Raspberry PI is not powerful enough to function as a miner, so two Raspberry Pi worked as nodes to join the private blockchain without mining. All devices use Go-Ethereum \cite{goethereum} as the client application to work on the blockchain network.

%------------------------------ Performance Evaluation -------------------------------------
\subsection{Performance Evaluation}
In the test scenario, two Raspberry Pi 3 devices are adopted to play the roles of the client and the service provider respectively. To measure the general cost incurred by the proposed BlendCAC scheme both on the IoT devices' processing time and the network communication delay, 50 test runs have been conducted based on the proposed test scenario, where the client sends a data query request to the server for an access permission. This test scenario is based on an assumption that the subject has a valid capability token when it performs the action. Therefore, all steps of authorization validation must be processed on the server side so that the maximum latency value is computed.

%Fig3_CapAc_exec_time
\begin{figure} [t]
\begin{center}
\begin{tabular}{c} 
\includegraphics[height=4.6cm]{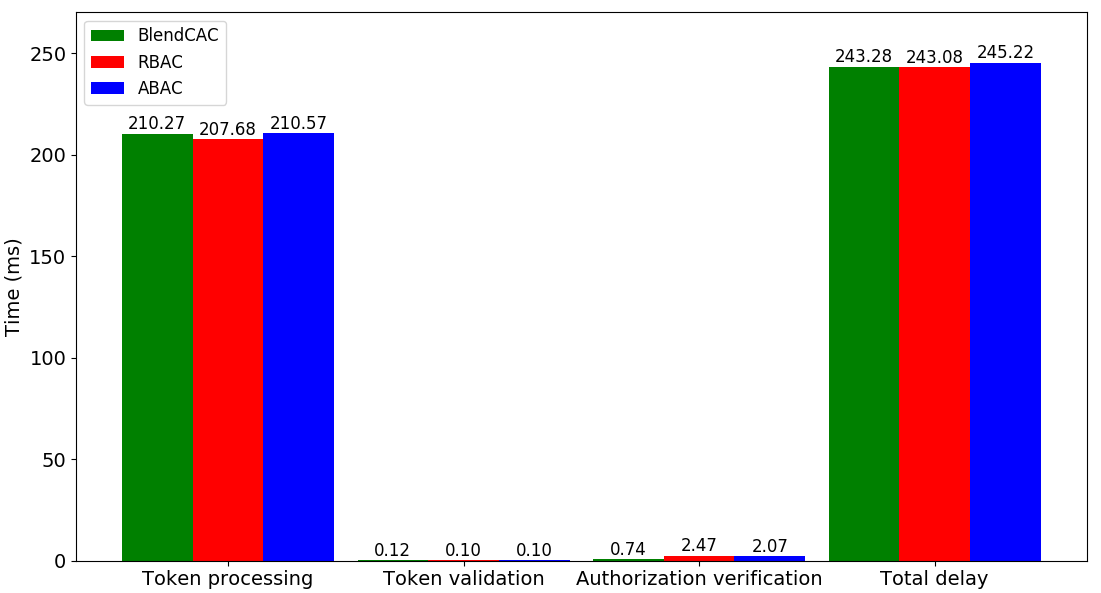}
\end{tabular}
\end{center}
\caption[example] { \label{fig:3-CapAc_exec_time} Computation Time for Each Stage in BlendCAC.}
\end{figure}

%\begin{enumerate}
%\item 
\subsubsection{Computational Overhead} 
According to the results shown in Fig. \ref{fig:3-CapAc_exec_time}, the average total delay time required by the BlendCAC operation of retrieving data from the client to server is 243 ms, which is almost the same as RBAC or ABAC does. The total delay includes the round trip time (RTT), time for querying capability data from the smart contract, time for parsing JSON data from the request, and time for access right validation. The token processing task is mainly responsible for fetching token data from the smart contract and introduces the highest workload among the authorization operation stages. Owing to the fact that encapsulating user-role relationship in the smart contract requires less data than the capability or attributes does, the RBAC incurred less computational cost than what BlendCAC and ABAC did in token processing stage. As the most computing intensive stage, the execution time of token processing is about 210 ms, which is accounted for almost 86\% of the entire process time. 

The entire authorization process is divided into two steps, token validation and authorization verification, where the average time of the authorization process is about 0.86 ms (0.12 ms + 0.74 ms). Compared with token validation process, which only simply checks the token valid status, the authorization verification process requires more computational power to enforce the local access control policies. Although the similarity in token data structure allows all the three access control models have almost the same time in the token validation stage, the BlendCAC outperforms the RBAC and ABAC in authorization verification. Since both the RBAC and ABAC needs a database to either manage user-role-permission relationship or maintain attributes-permission rules, which inevitably incurs time consuming on searching rules in database. In our experimental study, the RBAC (2.47 ms) and ABAC (2.07 ms) have much higher processing time than BlendCAC (0.74 ms).  

%\item 
\subsubsection{Communication Overhead}
Owing to the high overhead introduced by querying token data from the smart contract in token processing stage, a token data caching solution is introduced in the BlendCAC system to reduce network latency. When the client sends a service request to the server, the service side extracts cached token data from the local storage to valid authorization. The service providers regularly update cached token data by checking smart contract status. The token synchronization time is in consistence with block generation time, which is about 15 seconds in the Ethereum blockchain network. Simulating a regular service request allows us to measure how long it takes for the client to send a request and retrieve the data from the server. 

\begin{figure} [t]
\begin{center}
\begin{tabular}{c} 
\includegraphics[height=4.6cm]{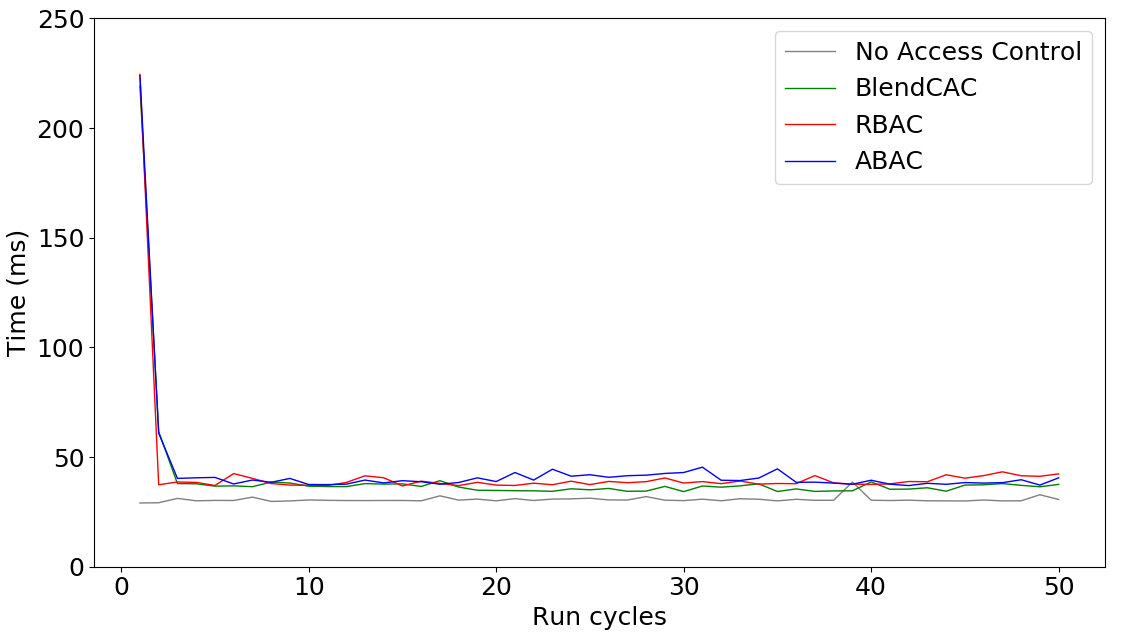}
\end{tabular}
\end{center}
\caption[example] { \label{fig:4-CapACVsNoCapAC} Network Latency of BlendCAC.}
\vspace{-10pt}
\end{figure}

Figure \ref{fig:4-CapACVsNoCapAC} shows the overall network latency incurred and compares the execution time of the BlendCAC with RBAC, ABAC and a benchmark without any access control enforcement. At the beginning, a long delay is observed in the first service request scenario, in which service provider communicated with the smart contract and cached the token data. However, through processing the local cached token data for authorization validation, the network latency decreases quickly and becomes stable at a low level during the subsequent service requests. The benchmark without access control enforcement takes an average of 31 ms for fetching requested data versus that the BlendCAC consumes on average of 36 ms. It means that the proposed BlendCAC scheme only introduces about 5 ms extra latency. The overhead in terms of delay is trivial. As shown by Fig. \ref{fig:4-CapACVsNoCapAC}, the BlendCAC also has lower latency than RBAC and ABAC in most period of time. In addition, unlike RBAC and ABAC, which rely on the local policy database as intermediate to valid access right, encoding access right directly in capability token makes the BlendCAC more scalable and flexible in the large scale IoT networks.
%\end{enumerate}

%Fig4_CapACVsNoCapAC

As the experimental results show, the proposed BlendCAC scheme introduced a small amount of overhead, both at the network layer and the local device layer. To measure general network latency of inter-domain communication, HTTP is executed on the same testbed to simulate a regular transaction, like connects, sends a request and retrieves the reply. Compared with calculated average network latency that is about 300 ms, the trade-off in the proposed BlendCAC is acceptable for the network environments by only incurring 5 ms latency (no more than 2\%). In addition, the test scenarios are based on Raspberry Pi devices, which belongs to a type of simple board computer (SBC) with limited computation power and memory space. It is reasonable to expect a better performance when the BlendCAC scheme is implemented on more powerful smart devices, like a smart phone. Although synchronizing cached token data with the smart contract requires more computational resource, the transactions of querying smart contract status are regularly launched by the service providers in a separate service thread rather than being called in each service request, so that network overhead over service request communication is greatly reduced to improve QoS requirement.

%------------------------------ Discussion -------------------------------------
\subsection{Discussions}
The experimental results demonstrate that the proposed BlendCAC strategy is effective and efficient to protecting the IoT devices from unauthorized access requests. Compared to centralized CapAC access models, the BlendCAC scheme has the following advantages:
\begin{enumerate}
\item \emph{Decentralized authorization}: leveraging the blockchain technique, the proposed BlendCAC scheme allows domain owners to control their devices and resource without depending on a centralized third authority to establish the trust relationship with unknown nodes; instead, it could define a domain-specific access authorization policy, which is critical to the scalable, heterogeneous and dynamic IoT-based applications;

\item \emph{Edge computing driven intelligence}: thanks to the blockchain technology, the BlendCAC scheme provides a device driven access control framework that is suitable for the distributed nature of IoT environments. Through transferring power and intelligence from the centralized cloud server to the edge of network, the risk of performance bottleneck and the single point of failure are mitigated, and the smart things are capable of protecting their own resource and privacy by enforcing user-defined security mechanism;

\item \emph{Fine granularity}: enforcing access right validation on local service providers enables the smart devices to make decisions whether or not to grant access to certain services according to the local environmental conditions. Fine-grained access control with the less privilege access principle prevents privilege escalation, even if an attacker steals a capability token; and

\item \emph{Lightweight design}: compared to the XML-based language for access control, such as XACML, JSON is a lightweight technology that is suitable for the resource constrained platforms. Given the experimental results, our JSON based capability token structure introduces small overhead on the computation and network performance.
\end{enumerate}

\section{Conclusions}
\label{sec:conclusion}  % \label{} allows reference to this section
In this paper, we proposed a decentralized capability-based access control framework leveraging the smart contract and blockchain technology, called BlendCAC, to handle the challenges in access control strategies for IoTs. A concept-proof prototype has been built in a physical IoT network environment to verify the feasibility of the proposed BlendCAC. The BlendCAC model is transcoded to smart contracts and works on the private Ethereum blockchain network. The desktops and laptops serve as miners to maintain the sanctity of transactions recorded on the blockchain, while Raspberry PI devices act as edge computing nodes to access and to provide IoT-based services. Extensive experimental studies have been conducted and the results are encouraging. It validated that the BlendCAC scheme is able to efficiently and effectively enforce access control authorization and validation in a distributed, trustless IoT network. This work has demonstrated that our proposed BlendCAC framework is a promising approach to provide a scalable, fine-grained and lightweight access control for IoT networks.

While the reported work has shown a wonderful potential, there is still a long way to go to build a complete decentralized security solution for IoTs and edge computing. Deeper insights are expected. Part of our on-going effort is focused on further exploration of the blockchain based access control scheme in real-world applications. Taking the smart surveillance system as a case study, the proposed BlendCAC will be extended to protect network cameras and motion sensors in the novel urban surveillance platform we developed recently \cite{chen2017enabling,xu2017real}.

% References
\bibliographystyle{IEEEtranS} % makes bibtex use IEEEtran.bst
\bibliography{report} % bibliography data in report.bib

\end{document}